\documentclass[prd,aps,secnumarabic,superscriptaddress,floatfix,preprintnumbers]{revtex4}
\usepackage[dvips]{graphicx,color}
\usepackage{dcolumn}
\def\sc{\mbox{\rule[-5pt]{0pt}{16pt}}}

\def\sb{\mbox{\rule{0pt}{11pt}}}

\def\sw{\mbox{\rule{24pt}{0pt}}}
\def\al{\alpha}

\def\ga{\gamma}
\def\de{\delta}

\def\Up{\Upsilon}
\def\Dot{\!\cdot\!}

\begin{document}
\preprint{NSF-KITP-09-189} \vspace{12pt}
\title{Hyperfine splittings in the $b\bar{b}$ system}
\author{Stanley F. Radford}\email{sradford@brockport.edu}
\affiliation{Department of Physics, The College at Brockport, State
University of New York, Brockport, NY 14420 \\ and \\
Kavli Institute for Theoretical Physics, Santa Barbara, CA 93106}
\email{sradford@brockport.edu}
\author{Wayne W. Repko}
\affiliation{Department of Physics and Astronomy, Michigan State
University, East Lansing, MI 48824}\email{repko@pa.msu.edu}
\date{\today}
\begin{abstract} Recent measurements of the $\eta_b(1S)$, the ground state of the $b\bar{b}$ system, show the splitting between it and the $\Up(1S)$
to be  69.5$\pm$3.2 MeV, larger than lattice QCD and potential model
predictions, including recent calculations published by us. The
models seem unable to incorporate such a large hyperfine splitting
within the context of a consistent description of the energy
spectrum and decays. We investigate whether a perturbative treatment
of our potential model (described below) can lead to such a
consistent description, including the measured hyperfine splitting,
by not softening the delta function terms in the hyperfine
potential. With this modification, we calculate the 1S hyperfine
splitting to be 67.5 MeV, with little effect on the overall fit of
our model results.  We also present predictions for the 2S and 3S
hyperfine splittings.
\end{abstract}
\maketitle
\section{INTRODUCTION}
Recent measurements by the BABAR Collaboration \cite{babar1, babar2}
have located the ground state of the $b\bar{b}$ system, the
$\eta_b(1S)$, at a mass of 9390.8$\pm$3.2 MeV. This value has
recently been confirmed by the CLEO Collaboration\cite{cleo}. Thus,
the hyperfine splitting between the 1S states is 69.5$\pm$3.2 MeV.
This splitting is surprisingly large when compared with the
corresponding splitting in charmonium, given the mass dependence of
the leading order contribution. In addition, this hyperfine
splitting is larger than predictions in recent lattice QCD
calculations, which range from 52.5$\pm$1.5 MeV to 65.8$\pm$4.6 MeV
\cite{lqcd1,lqcd2,lqcd3,lqcd4}, as well as those predicted by recent
potential models (42$\pm$13 MeV) \cite{desl}.  A comparison of the
data with various model results is presented in the concluding
section of this paper.

This large hyperfine splitting has recently been investigated as a
possible indicator of new physics \cite{desl}, as a means of
adjusting the value of the strong coupling parameter \cite{css}, and
to fix the value of $\al_S$ in an investigation other quarkonium
states \cite{bbd}. While further data should clarify these points,
in this paper we investigate whether the model we employed in our
2007 paper \cite{rr1}, is robust enough to accommodate this new
data.

The results in \cite{rr1}, while providing a good overall fit to the
$b\bar{b}$ spectrum, do not yield such a large hyperfine splitting.
In that paper, we showed that the perturbative treatment of a model
consisting of a relativistic kinetic energy term, a linear confining
term including its scalar relativistic corrections and the complete
perturbative one-loop quantum chromodynamic short distance potential
was able to reproduce the overall spectrum of the $b\bar{b}$ system
as well as its radiative decays, with good accuracy. However, in our
model results, the 1S hyperfine splitting, at 47 MeV, was
considerably smaller than the BABAR measurements indicate.

In our previous calculation, we followed the standard practice of
softening the delta function terms in the potential \cite{gi}. This
is essential when the entire interaction is treated
non-perturbatively in order to avoid instability in the numerical
calculations. However, in a perturbative treatment, we can perform
the calculations with the delta function terms unsoftened and still
retain the overall goodness of our fit for the spectrum and leptonic
decays, while reproducing the correct 1S hyperfine splitting, with
only minor changes in the potential parameters.

\section{HEAVY QUARKONIUM HYPERFINE POTENTIAL}

The one-loop hyperfine terms arising in the QCD potential are
\cite{grpot,bnt}

\begin{equation}
V_{HF}=\frac{32\pi\al_S\vec{S}_1\Dot\vec{S}_2}{9m^2}\left\{\left[1-\frac{\al_S}{12\pi}
(26+9\ln\,2)\right]\de(\vec{r})-\frac{\al_S}{24\pi^2}(33-2n_f)\nabla^2\left[\frac{\ln\,\mu
r+\ga_E}{r}\right]+\frac{21\al_S}{16\pi^2}\nabla^2\left[\frac{\ln\,
mr+\ga_E}{r}\right]\right\},
\end{equation}
where $\al_S$ is the strong coupling constant in the GR scheme
\cite{gr1}.

A variety of approaches to softening the delta function terms have
been used \cite{bgs}.  In most instances, this softening is done
because the approach to integrating the Schr\"odinger equation
precludes the inclusion of delta function terms. In our previous
calculation we chose to adopt the quasistatic approximation
\cite{gjrs}, and softened the delta function as

\begin{equation}
\de(\vec{r})\to\frac{m^2}{\pi r}e^{-2mr}.
\end{equation}

In the interest of clarity, the reevaluation of the hyperfine splittings presented here begins with a variational calculation of the wave functions and energies of an unperturbed Hamiltonian. The Hamiltonian consists of the relativistic kinetic energy, a linear confining potential and a short-ranged Coulomb-like potential that includes the one-loop correction to the strong coupling parameter $\alpha_S$. This procedure results in an orthogonal set of radial wave functions for every orbital angular momentum. These radial functions and the associated energies describe the spin-averaged upsilon spectrum. We include the contributions of all $v^2/c^2$ and one-loop QCD corrections to the the $b\bar{b}$ potential energy using perturbation theory, retaining any delta function terms that arise in the derivation of these corrections. Further details can be found in Ref.\,\cite{rr1}.

The retention of the delta function terms in our calculation is motivated by their use in familiar QED applications. The inclusion of delta function contributions in perturbative calculations is necessary to reconcile the hydrogen fine-structure results with the Dirac equation, to derive the hydrogen hyperfine splitting, and to understand the positronium and muonuim spectra \cite{grs}. Since the wave functions in Ref.\,\cite{rr1} are derived variationally using a non-singular Hamiltonian, there are no technical difficulties in including delta function terms perturbatively. Given the numerous approaches to the calculation of the hyperfine splittings in quarkonia, including the delta function approach of Ref.\cite{bbd}, our perturbative evaluation of the hyperfine intervals is relevant to the overall discussion.

\section{RESULTS}

As discussed above, we have recalculated the $b\bar{b}$ spectrum by
retaining the delta function terms in our perturbative calculation.
The resulting fitted parameters are shown in Table \ref{params},
along with those from Ref. \cite{rr1}.

\begin{table}[h]
\centering
\begin{tabular}{ldddd}
\toprule
  &\multicolumn{1}{r}{\sb Softened \protect\cite{rr1}}\sw &\multicolumn{1}{r}{\sb Unsoftened}\sw\\
\hline
\sc$A$ (GeV$^2$) & 0.177_{-0.002}^{+0.006}    & 0.175 \\
\hline
\sc $\al_S$      & 0.296_{-0.007}^{+0.004}    & 0.295 \\
\hline
\sc $m_q$ (GeV)  & 5.36_{-0.42}^{+0.87}     & 5.33  \\
\hline
\sc $\mu$ (GeV)  & 4.74     & 4.82  \\
\hline
\sc $f_V$        & 0.00    & 0.00  \\
\botrule
\end{tabular}
\caption{Fitted parameters for the softened and unsoftened
potentials\label{params}}
\end{table}
As can be seen from Table \ref{params}, the retention of the delta
function terms leads to a very minor adjustment of the fitted
parameters. Our results for the $b\bar{b}$ spectrum are shown in
Table \ref{bottomspec}.  It can been seen that the only significant
changes are in the s-state hyperfine splittings.  We calculate these
splittings to be: 67.5 MeV for 1S, 35.9 MeV for 2S, 30.3 MeV for 3S.
The latter two values are, of course, predictions.

\begin{table}[h]\centering
\begin{tabular}{lddd} \toprule
\multicolumn{1}{c}{\sc}$m_{b\bar{b}}$\,(MeV)  &\multicolumn{1}{c}{Softened}  & \multicolumn{1}{c}{ Unsoftened}& \multicolumn{1}{c}{ Expt } \\
\hline
\sb$\eta_b(1S)$\mbox{\rule{12pt}{0pt}}   & 9413.70\sw\sw   & 9392.91  & 9390.8\pm 3.2 \\
\hline
\sb$\Up(1S)^*$         & 9460.69   & 9460.38  & 9460.30\pm 0.26  \\
\hline
\sb$\chi_{b\,0}(1P)^*$ & 9861.12   & 9861.39  & 9859.44\pm 0.52  \\
\hline
\sb$\chi_{b\,1}(1P)^*$ & 9891.33   & 9891.33  & 9892.78\pm 0.40  \\
\hline
\sb$\chi_{b\,2}(1P)^*$ & 9911.79   & 9910.63  & 9912.21\pm 0.40  \\
\hline
\sb$h_b(1P)$      & 9899.99   & 9899.93    &   \\
\hline
\sb$\eta_b(2S)$   & 9998.69   & 9987.42    &   \\
\hline
\sb$\Up(2S)^*$    & 10022.5   & 10023.3    & 10023.26\pm 0.31 \\
\hline
\sb$\Up(1D)$      & 10149.5   & 10149.8    &    \\
\hline
\sb$1^3D_2$       & 10157.1   & 10157.3    & 10161.1\pm 1.7    \\
\hline
\sb$1^3D_3$       & 10162.9   & 10163.1    &    \\
\hline
\sb$1^1D_2$       & 10158.4   & 10158.6    &    \\
\hline
\sb$\chi_{b\,0}(2P)^*$ & 10230.5   & 10230.5    & 10232.5\pm 0.6  \\
\hline
\sb$\chi_{b\,1}(2P)^*$ & 10255.0   & 10254.8    & 10255.46\pm 0.55\\
\hline
\sb$\chi_{b\,2}(2P)^*$ & 10271.5   & 10271.2    & 10268.65\pm 0.55\\
\hline
\sb$h_b(2P)$    & 10262.0   & 10261.8    &                  \\
\hline
\sb$1^3F_2$     & 10353.0   & 10353.1    &                 \\
\hline
\sb$1^3F_3$     & 10355.8   & 10355.8    &                 \\
\hline
\sb$1^3F_4$     & 10357.5   & 10357.5    &                 \\
\hline
\sb$1^1F_3$     & 10355.9   & 10356.0    &                 \\
\hline
\sb$\eta_b(3S)$ & 10344.8   & 10333.9    &                  \\
\hline
\sb$\Up(3S)$    & 10363.6   & 10364.2    & 10355.2\pm 0.5   \\
\hline
\sb$\Up(2D)$    & 10443.1   & 10443.0    &                  \\
\hline
\sb$2^3D_2$     & 10450.3   & 10450.1    &                  \\
\hline
\sb$2^3D_3$     & 10455.9   & 10455.7    &                  \\
\hline
\sb$2^1D_2$     & 10451.6   & 10451.4    &                  \\
\hline
\sb$2^3F_2$     & 10610.0   & 10609.6    &                 \\
\hline
\sb$2^3F_3$     & 10613.0   & 10612.5    &                 \\
\hline
\sb$2^3F_4$     & 10615.0   & 10614.5    &                 \\
\hline
\sb$2^1F_3$     & 10613.2   & 10612.7    &                 \\
\hline
\sb$\eta_b(4S)$ & 10622.8   & 10609.4    &                  \\
\hline
\sb$\Up(4S)$    & 10643.0   & 10636.4    & 10579.4\pm 1.2   \\
\botrule
\end{tabular}
\caption{Results for the $b\bar{b}$ spectrum using softened and
unsoftened potentials are shown. Our perturbative fits use the indicated
states. The value of the $\eta_b(1S)$ mass is taken from \protect\cite{babar2} and all other data is taken from \protect\cite{pdg}.}\label{bottomspec}
\end{table}
We have also examined the leptonic widths as shown in in Table
\ref{botleptonic}.  We find that there is a noticeable increase only
for the 3S and 4S states. However, the modified results are still
compatible with the experiments.
\begin{table}[h]
\centering
\begin{tabular}{lddd} \toprule
\multicolumn{1}{c}{\sc $\Gamma_{e\bar{e}}$\,(keV)}  &\multicolumn{1}{c}{Softened}  &\multicolumn{1}{c}{ Unsoftened}& \multicolumn{1}{c}{\,\, Expt}   \\
\hline
\sb$\Up(1S)$\mbox{\rule{12pt}{0pt}} & 1.33\sw\sw  & 1.33  & 1.340\pm 0.018 \\
\hline
\sb$\Up(2S)$   & 0.61  & 0.62  & 0.612\pm 0.011 \\
\hline
\sb$\Up(3S)$   & 0.46  & 0.48  & 0.443\pm 0.008 \\
\hline
\sb$\Up(4S)$   & 0.35  & 0.40  & 0.272\pm 0.029 \\
\botrule
\end{tabular}
\caption{The leptonic widths of the $\Up(nS)$ states using softened
and unsoftened potentials.}\label{botleptonic}
\end{table}
\section{CONCLUSION}

We conclude with a comparison of various modelling approaches with
the experimental data.  In Table \ref{hfcompare}, we show a
comparison of the data with the results of representative model
calculations: our new (unsoftened) and old (softened) models,
several recent lattice QCD results, next-to-leading logarithmic
perturbative QCD, and a QCD-inspired phenomenological treatment.
This comparison shows the range of predictions of various modelling
approaches. Additional data should provide clarification.
\begin{table}[h]
\centering
\begin{tabular}{c|c|c|c|c|c|c|c|c|c}
\toprule
\multicolumn{1}{c}{\sc $\Delta HF$\,(MeV)}  &\multicolumn{1}{|c}{\,\,Our\,\,}  &\multicolumn{1}{|c}{Ref.\,\cite{rr1}}  &\multicolumn{1}{|c}{Ref.\,\cite{lqcd1}}  &\multicolumn{1}{|c}{Ref.\,\cite{lqcd2}}  &\multicolumn{1}{|c}{Ref.\,\cite{lqcd3}}  &\multicolumn{1}{|c}{Ref.\,\cite{lqcd4}}  &\multicolumn{1}{|c}{Ref.\,\cite{nllk}}  &\multicolumn{1}{|c}{Ref.\,\cite{bbd}}  &\multicolumn{1}{|c}{Expt}\cite{babar2}   \\
\hline
\sb$\Delta HF(1S)$   & 67.5$\pm$0.7  & 47.0  & 52.5$\pm$1.5  & 54.0$\pm$12.4  & 61$\pm$14  & 65.8$\pm$4.6  & 39.5$\pm$8  & 63.4  & 69.5$\pm$ 3.2 \\
\hline
\sb$\Delta HF(2S)$   & 35.9$\pm$0.3  & 23.8  &  &  & 30$\pm$19  &  &  & 35.0  &          \\
\hline
\sb$\Delta HF(3S)$   & 30.3$\pm$0.2  & 18.8  &  &  &  &  &  & 27.6  &           \\
\botrule
\end{tabular}
\caption{Comparison of s-state hyperfine splittings in various models with experiment. The errors on the hyperfine splittings in column `Our' were obtained by using the Gaussian errors on the parameters $\alpha_S$, $m_b$ and $\mu$ associated with our fit to the upsilon spectrum. }\label{hfcompare}
\end{table}

We have shown that by not softening the delta function terms which
arise in the one-loop quark-antiquark potential, we are able to
reproduce the surprisingly large hyperfine splitting in the 1S level
of the $b\bar{b}$ system. This is in contrast to other modeling
results, including ours, which soften the delta function terms. In
addition, the improvement in the hyperfine splittings does not
affect either the overall fit of our perturbative calculation to the
data, nor to our ability to reproduce the leptonic decay widths. We
also predict somewhat larger hyperfine splittings for the other
s-states.

\begin{acknowledgments}
SFR thanks the Kavli Institute for Theoretical Physics for its
hospitality during June 2009 and July 2010. This research was
supported in part by the National Science Foundation under Grants
PHY-0555544 and PHY-0551164.
\end{acknowledgments}

\end{document}